\def\be#1{\begin{equation}\label{eq:#1}}
\def\ee{\end{equation}}
\def\EC#1{(\ref{eq:#1})}

\documentclass[12pt,preprint]{aastex}

\slugcomment{ }

\shorttitle{HI in NGC~2613}
\shortauthors{Irwin\& Chaves}

\begin{document}


\title{High Latitude HI in NGC~2613: Buoyant Disk-Halo Outflow}


\author{Judith A. Irwin \& Tara Chaves}
\affil{Dept. of Physics, Queen's University,
    Kingston, Canada, K7L 3N6}
\email{irwin@astro.queensu.ca, tchaves@astro.queensu.ca}




\begin{abstract}
We combine new VLA D array HI data of NGC~2613 with previous high resolution
data to show new disk-halo features in this galaxy.
The
global HI distribution is modeled in detail using a technique
which can disentangle the effects of inclination from scale height
and can also solve for the average volume density distribution in
and perpendicular to the
disk.  The model shows that the galaxy's
inclination is on the low end of the range given by Chaves \& Irwin (2001)
and that the HI disk is thin ($z_e$ = 188 pc), showing no evidence for halo.    

Numerous discrete disk-halo features are observed,
however, achieving
$z$ heights up to 28 kpc from mid-plane.  One
prominent feature in particular,
 of mass, $8\,\times\,10^7$ M$_\odot$ and height, 22 kpc,
is seen on the advancing side of the galaxy
 at a projected galactocentric radius of 15.5 kpc.  If
this feature achieves such high latitudes because of events in the disk
alone, then input energies of order $\sim$ 10$^{56}$ ergs are required.
We have instead investigated the feasibility of such a large feature
being produced
via buoyancy (with drag) within a hot, pre-existing X-ray
corona.
Reasonable 
plume densities, temperatures, stall height ($\sim$ 11 kpc),
outflow velocities and ages 
can indeed be achieved in this way.
The advantage of this scenario is that the input energy need only
be sufficient to produce blow-out, a condition which requires
a reduction of three orders of magnitude in energy.  If this is
correct, there should be an observable X-ray halo around
NGC~2613.

\end{abstract}




\section{Introduction}
\label{introduction}

High latitude HI in disk galaxies is increasingly being recognized
as an important tracer of disk activity and dynamics.
 The gas scale height is a measure of time integrated
 activity in the disk and discrete features trace the
level of local activity.
The disk-halo interface represents a severe transition between 
two very different
environments:  the high density, low velocity dispersion disk and the 
low density, high velocity dispersion halo.  Globally,
it may provide information
on the shape of the dark matter halo (e.g. Olling \& Merrifield 2000).
In a few recent cases,
lagging HI
 halos have also been detected, such that the high latitude gas is
rotating
$\sim$ 20 - 50 km s$^{-1}$ more slowly
than the underlying disk
(Schaap, Sancisi, \& Swaters 2000, Fraternali et al. 2001,
Lee et al. 2001, Rand 2000 and T{\"u}llmann et al. 2000).
Explanations for such lags
have focussed mainly on galactic
fountains (cf. Bregman 1980, Swaters et al.
1997, Schaap et al. 2000,
Collins et al. 2002).  Edge-on galaxies are particularly
useful in the study of the disk-halo region and,
in this paper, we especially wish to investigate how
discrete disk-halo features
 can be used as probes of
the environment into which they are emerging.

The edge-on galaxy, NGC2613 (D = 25.9 Mpc)
is a good candidate for
such a study.  In Paper I (Chaves \& Irwin 2001),
we presented Very Large Array (VLA) C Array observations
of this galaxy.  The resulting C array 
zeroth moment map over an optical image 
is shown in Fig.~\ref{Carray}, illustrating
the presence of six symmetrically placed vertical extensions in this galaxy.
In this paper, we present new VLA D array HI data of NGC~2613 and
examine them separately as well as in combination with the previously
obtained C array data.  The observations are presented in 
Sect.~\ref{observations}, the results, including a description of
extremely large high latitude HI features and
a model of the global HI distribution are
 given in
Sect.~\ref{results}, in Sect.~\ref{discussion},
we present the discussion, including a feasibility study of 
a buoyancy model, and a summary is given in Sect.~\ref{conclusions}.

\section{Observations}
\label{observations}

Details of the C array data acquisition and mapping have been given in Paper I.

The D array data were taken on Sept. 15, 2000.  The flux calibrator was
3C~48 and the phase calibrator was 0837-198.  The flux calibrator was
also used as a bandpass calibrator and was observed once for 17 minutes
at the beginning of the observing period. 
The phase calibrator
was observed in 3 minute scans alternating with 25 minute scans on
NGC~2613. The total on-source observing time was 2.4 hours and 
on-line Hanning velocity smoothing was applied.  The data were edited,
calibrated, and
Fourier Transformed using natural weighting, cleaned,
and continuum subtracted in the usual way using standard programs in
the Astronomical Image Processing System (AIPS) of the National Radio
Astronomy Observatory.  The resulting (RA, DEC, Velocity) cube was
then examined for broad scale structure.

The calibrated D array UV data were then combined with the
previously obtained C array UV data (on-source observing time =
15.3 hours, see Paper I)
to improve the signal to noise (S/N) ratio.  Since the high resolution
C array data have already been presented, we here concentrate on the
broader scale structure by re-mapping the C+D array data,
using natural weighting, over a UV range corresponding to the D array
data only, i.e. 5 k$\lambda$.  That is, we show the lower resolution
results (at a typical D array resolution), 
but with the highest S/N possible.
Table 1 lists observing and map parameters
for D array and for the combined C+D array data.  Note that,
although we show images for the combined data set, we have
inspected and analyzed
all data sets both independently and in combination.  

\section{Results}
\label{results}

Note that a full description of the C array results and relevant
images have been
given in Paper I and we have reproduced the zeroth moment C array map
in Fig.~\ref{Carray}.  The remaining 
The zeroth and first moment maps from the combined data set are shown in 
Fig.~\ref{moments},
both rotated so that the x axis is parallel to the major axis of the galaxy.
Vertical slices have been taken perpendicular to the major axis at positions
numbered 1 through 10 in Fig.~\ref{moments}a 
and these are shown 
as position-velocity (PV) plots in Fig.~\ref{PV}.  
The PV slices in the last two panels 
of Fig.~\ref{PV} show
averages over  the 
receding and advancing sides of the galaxy.

\subsection{Tidal Tail}

For the first time, we see that there is a tail
of emission on the south-eastern
(left, Fig.~\ref{moments}a)
 side of the galaxy which resembles tidal tails seen in
galaxy interaction simulations.  This tail is
likely produced via an interaction with the
companion, ESO~495-G017, to the north-west, whose
systemic velocity is separated from that of NGC~2613
by only 143 km s$^{-1}$.
 The tail is at negative velocities
with respect to the receding side of the galaxy from which it originates
and shows a strong velocity gradient along its length,
becoming negative with respect to the galaxy's systemic velocity
as well, i.e. the velocities become forbidden.  This can
occur in the case of interactions but is difficult
to explain any other way.   Given this velocity structure,
the tail must be trailing from
the eastern, receding side of NGC~2613 and is
 therefore in front of the galaxy.
The anomalous emission seen 
around -200 km s$^{-1}$ in Slice 2, as well as
the average over the receding side of the galaxy (second last panel)
 of Fig.~\ref{PV} belongs to this tail as well.  There is
excess emission on the north-west (right, Fig.~\ref{moments}a)
side of the galaxy 
 in the vicinity
of the companion, but we have found no systematic velocities
associated with that emission or any other sign of a corresponding redshifted
feature.

\subsection{High Latitude HI}
\label{disk-halo}

In Paper I, 
using the C array data alone, we identified and characterized the
properties of 6, symmetrically placed extensions
labelled 
F1 to F6 in Fig.~\ref{Carray}.  Fig.~\ref{moments}a, which
includes the
lower D array resolution data, now
also reveal these same features on broader scales,
some of which are blended with other features. 
Since
a single cutoff level must be applied over the whole
galaxy to create this image, 
 this moment
map is best used to see a global view of the disk-halo features
and identify 
where new features might be present.  The full
{\it z} extents and, of course, the velocities
of the individual features are best measured
 from PV slices perpendicular to the major axis.
We have
inspected many PV slices over the map
and show selected slices
 (labelled
1 through 10 in Fig.~\ref{moments}a) in 
Fig.~\ref{PV}. 
  In the new data, 
the previously seen
features F1 through F6 occur at the same velocities and 
the same positions 
as shown in Fig.~\ref{Carray},
after taking into account the different
beam sizes and the addition of broader scale 
emission.  Please refer to
both Figs.~\ref{moments}a and \ref{PV} for a more thorough
discussion of these features, presented below.

F1 appears similar to what has 
previously been seen  (Fig.~\ref{Carray}), 
but at lower resolution.
The PV plot for F1 can be seen in Slice 2 
which reveals it as a distinct feature at a velocity
of 250 km s$^{-1}$ but it can also be seen in Slice 1
as well,  extending to 
a projected {\it z} height of
at least 100$^{\prime\prime}$ (12.6 kpc)
``above'' (i.e. positive {\it z}) mid-plane.
A newly seen
 extension immediately to the west of F1, sampled by Slice
3, may actually be associated with or part of F1, since
it is seen in PV space as a small arc at about the same
velocity as F1.
 
F2 (see Slice 1) 
shows several velocity components and 
is more complex 
than previously seen. Like F1, F2 extends to at least
100$^{\prime\prime}$ at a velocity of 210 km s$^{-1}$.
It is not clear whether the other components may be
blended with emission from the tidal tail.
 
F3 reveals more structure than the extension in Fig.~\ref{Carray}
and is sampled by Slice 4 
which reveals it as a 2-pronged arc above
the disk centered at a velocity of 100 km s$^{-1}$.
Such double peaks in velocity space are typical of
such features and have been seen in other galaxies
(see, e.g. Lee \& Irwin 1997, Lee et al. 2001).   

The extension to the left of the F4 label in
Fig.~\ref{moments}a) extends to a height 225$^{\prime\prime}$
(28 kpc)
below (i.e. negative {\it z}) mid-plane in this
map.  As can be
seen in Slice 3 of the PV map, this feature occurs at
a velocity of 50 km s$^{-1}$ and shows no velocity gradient
with height.   
 F4 reaches the highest {\it z} of all
the extensions and, 
to our knowledge, 
 reaches higher
latitudes than any previously seen in an edge-on galaxy,
exceeding even the starburst galaxy, NGC~5775, whose 
known HI features extend
to $\sim$ 7 kpc from the plane (Lee et al. 2001).
  This is
remarkable, given that NGC~2613
(log L$_{FIR}$ = 10.00) has a massive (0.1 $\to$ 100 M$_\odot$)
star formation rate
that is a factor of 3 lower than that of NGC~5775 and a
supernova input energy rate per unit area that is a factor of
7 less than NGC~5775
(Irwin et al. 1999).
  
Feature F5 is sampled by Slices 8 and 9 and can most clearly
be seen in Slice 9 as a vertical feature above the plane
in PV space at a velocity
of -300 km s$^{-1}$.  Whereas it 
appeared as a single extension in Fig.~\ref{Carray}, it
  can now be
seen to have a counterpart 55$^{\prime\prime}$ (7 kpc)
in projection
farther out along the disk which is sampled by Slice
10.  The counterpart clearly occurs 
at the same velocity and also shows no
velocity gradient with height.  These two features may
form the base of a large loop
or cylinder, similar to what has been
seen in galaxies like NGC~5775 (Lee et al. 2001). 

The most interesting and well-defined feature, however,
is associated with F6.   For example, slices
6, 7,  8 and 9 pass through
a feature below 
 the major axis which extends 22 kpc (175$^{\prime\prime}$)
in projection, from the plane
and centers at -307 km s$^{-1}$. 
Individual channel maps 
of this feature are shown in Fig.~\ref{channels}.
 We will refer to this as the -307 km s$^{-1}$
feature, rather than F6 since, as revealed by the
new data, the feature is more extensive than what was 
originally seen in Fig.~\ref{Carray}.
This feature will be discussed  extensively
 in  Sect.~\ref{307feature}.

This disk-halo emission is faint (3 - 4 $\sigma$ typically,
note contour spacings in Fig.~\ref{PV}) but we consider 
the above-discussed features to be real because,
{\it 1)} their
spatial morphology is typical of the kinds of extensions
and loops seen in other edge-on systems, 
{\it 2)} even though the
feature might be faint far from the disk, there is
evidence for disturbed contours `beneath them'
even at highly significant contour
levels (see the 5th and 6th contours
of Fig.~\ref{moments}a, for example), 
{\it 3)} at least the 6
major features, F1 to F6 can be seen in both independent
data sets, 
{\it 4)} the features generally occur over several
independent beams and/or
velocity channels, and
{\it 5)} in PV space, the features are distinct, yet
smooth and connect continuously to the disk, unlike
artifacts produced by a badly cleaned beam 
(see, e.g. the arc above the disk
seen in Slice 3 of Fig.~\ref{PV} 
at $\sim$ 270 km s$^{-1}$ or the larger structures in
Slices 9 and 10). 
In addition, {\it 6)} 
the Receding and Advancing panels of 
panels of Fig.~\ref{PV}) show {\it averages} over the
entire advancing and receding sides of the disk, rather than
just selected individual slices as shown elsewhere
in the figure.  In these panels, therefore, we have basically
averaged together signal with noise and weak extensions
would be diluted (i.e. ``beam-averaged'', where the beam
is the size of half the galaxy).  Yet these panels still
show the extensions quite clearly, especially on the Advancing
side where the disk-halo emission is dominated almost
entirely by the -307 km s$^{-1}$ feature.  This feature
is seen up to the 4th contour which is a 6$\sigma$
detection in the diluted averaged panel.
Finally,
{\it 7)} for at least 4 of the 6 features, there is 
evidence from independent observations that extraplanar emission
is present. For example, our radio continuum image
(see Fig. 9 of Paper I) shows
extended emission at the positions of F3, F4, F5, and F6.
Also, an early HI image by Bottema (1989)
at lower resolution shows extended
HI emission at roughly the same locations.

The zeroth moment map of Fig.~\ref{moments}a also shows a number of 
low intensity ``disconnected"
features at high latitudes (for example, the emission around
 -300$^{\prime\prime}$, -200$^{\prime\prime}$).
We do {\it not} claim that all of these are real, but we have
chosen to display them for the sake of comparison with
future observations and because some of the emission shown in
independent lower resolution images (e.g. Bottema 1989, Irwin et al. 1999)
is so extended.  An
 example would be the `knot' of
emission between F3 and F5 which is sampled by Slice 6.  For
this feature, the velocities are discordent with respect to the
underlying disk (see Fig.~\ref{PV}) and therefore do not
satisfy the criteria listed above.

Particularly interesting about these very high latitude features is
that there is little or no evidence for any lag in velocity with {\it z},
even to very high distances from the place.
The features seen below the plane in slices 3, 8, 
and 9 and features seen above the plane in slices 9 and 10
show no velocity gradient. The early data of Bottema (1989) 
also showed that the high latitude HI was co-rotating.
 There is some evidence, in our data, for curvature in
the smaller features (cf. Slices 3 \& 4 
above the plane), but 
one could not make any case for a 
lag.  The only slice which shows
some evidence for lagging gas is Slice 1, but this slice is 
also closest to
the tidal tail, confusing the interpretation.  The averages
of the advancing and receding sides of the galaxy also do
not support the presence of a lagging halo.  Here, we see an
asymmetry in the sense of
many disk-halo features on the receding side, but the advancing
side is dominated by the single -307 km s$^{-1}$ feature. 
None of the features, however, show a decline with {\it z}.
  We investigate
this
further in the next section.

\subsection{Kinematic Models of the Global HI Distribution}
\label{models}

We have modeled the
 global HI density and velocity distributions for NGC~2613 
 following Irwin \& Seaquist (1991) and Irwin (1994).
This approach models all spectra in the HI cube by adopting a volume
density distribution in and perpendicular to the plane, a rotation curve
and velocity dispersion (if
required), and a position and  orientation on the sky.  Given 
trial input 
parameters describing
these distributions, the routine then varies the parameters, examining the
residuals, until the best fit solution is found.  Various sections of
the galaxy can be modeled, including 
galactocentric rings in the galaxy's plane.  In the current 
version, we
can also isolate
slices parallel to the galaxy's plane.  This permits modeling the
halo, independent of the disk, which is important in this study
in the event that a lagging HI
halo might be present. While there is no
evidence for lags in the discrete features shown in
 Fig.~\ref{PV}, it is possible
that the smooth, low intensity halo  might be better fit
with a lagging velocity distribution, similar to what has been found
for NGC~2403 (Fraternali et al. 2001).
The routine, called CUBIT, interfaces to classic AIPS;
potential users are requested to contact the first author for the code.

For NGC~2613, the 
adopted in-plane density distribution is a gaussian ring, described by:
$n(R,z=0)\,=\,n_0 \,exp({(R - R_0)}^2/D^2)$, where $R$ is the galactocentric
radius, $R_0$ is the galactocentric distance of the center of the ring,
$n_0$ is the in-plane density at the center of the ring, and $D$ is
the scale length.  $D$ can have different values for points $R>R_0$ 
($D_o$) and for points $R<R_0$ ($D_i$).  The distribution perpendicular
to the plane is given by the exponential, 
$n(R,z)\,=\, n(R,z=0)\,exp(-z/H_e)$, where $z$ is the perpendicular
distance from midplane and $H_e$ is the exponential scale height.
The rotation curve is described by the Brandt curve,
$V(R)\,=\, V_{sys} \,\pm\,V_{max}\,(R/R_{max})\,
{(1/3\,+\,2/3\,{(R/R_{max})}^m)}^{(-3/2m)}$, where
$V_{sys}$ is the galaxy's systemic velocity, 
$V_{max}$ is the peak of the rotation curve occurring at a galactocentric
radius of $R_{max}$ and $m$ is the Brandt index which is a shape
parameter.
 A gaussian velocity dispersion,
$\sigma_V$ can also be applied, if needed.  
Note that a velocity dispersion will result from
all contributions to line widening along the entire   
 line-of-sight
direction, for example,  any
non-circular motions along the plane, streaming motions
due to spiral arms and local turbulence.  Thus, $\sigma_V$ 
may be larger than what might be expected from a local value.
 The above distributions result
in 10 free parameters:
 $n_0$, $R_0$, $D_o$, $D_i$, $H_e$, $V_{sys}$, $V_{max}$, 
$R_{max}$,  $m$,
and $\sigma_V$. In addition, there are 4 more
orientation parameters:  the position of the galaxy center, RA(0), DEC(0), 
the major axis position angle, PA, and the inclination, i.  
Over the region
of the cube occupied by NGC~2613, there are 552 independent
data points (beam / velocity-channel resolution elements) 
above a 2 $\sigma$ intensity level in the
combined C+D cube, so the 
14 parameters are well constrained, over all. 
Certain parameters may show a larger variation, however, depending
on the nature of the distribution;
for example, $R_{max}$ tends to have a larger error 
since the point at which the rotation curve changes from rising
to flat is not as well constrained by the data 
as some of the other parameters.

Table 2 lists the best fit modeled values for NGC~2613 for the combined
C+D array data.  
We also modeled the C array data and D array data alone;
the error bars of Table 2 reflect the variations in the
parameters over the different arrays 
that result.  Given the range of resolution over these arrays,
from 26.1$^{\prime\prime}$ $\times$ 20.2$^{\prime\prime}$ 
(3.3 $\times$ 2.5 kpc)
to 
96.5$^{\prime\prime}$ $\times$ 42.9$^{\prime\prime}$
(12.1 $\times$ 5.4 kpc)
the error bars on the modeled parameters are very small.
The best fit model is shown as dashed curves superimposed on the data
in Fig.~\ref{PV}.  
Taking only residuals (data cube - best fit model cube)
greater than 3$\sigma$, the root-mean-square value
is 1.77 mJy beam$^{-1}$ and the average relative error is $\sim$ 10\%.
A zeroth moment map made from
the cube of the residuals of the best fit is shown in 
Fig.~\ref{residuals}.

To determine whether there are  changes in i, PA, and $H_e$
with galactocentric radius, R, we also modeled the galaxy in concentric rings.
For example, either a lower value of i with increasing R (a warp) 
or a higher value of $H_e$ with
increasing R (i.e. a flare) could conspire to make the halo of a 
galaxy appear to lag.  We find that the inclination does not decrease
with radius (there is a slight increase).  There is indeed evidence
for flaring such that, in each data set, $H_e$ systematically increases
with R.  The most significant change is measured for the D array data alone such
that the mean scale height between 
a radius of 140$^{\prime\prime}$ and 300$^{\prime\prime}$
(17.6 to 38 kpc)
is 5.5$^{\prime\prime}$ (693 pc) compared to 
a value of 1.7$^{\prime\prime}$ (214 pc) over the whole galaxy.
Thus, the disk is quite thin with a modest flare at large
galactocentric radii.

In order to model the halo alone, a variety of 
lower and upper cutoffs to {\it z} were 
applied and these data were modeled in a similar fashion. 
(Note that {\it z} refers to the actual distance from the
galaxy's midplane, not just the
projected distance from the major axis.)
 The number of
independent data points is reduced typically by about a factor of
3, depending on which {\it z} limits are chosen, but
we have also held some of the parameters
(the central position, position angle and systemic velocity) 
 fixed.  For the halo models, we find {\it no} evidence that
$V_{max}$ could be lower at higher {\it z}. The same conclusion is reached
if the C array and D array data are modeled independently.
The absence of a lagging HI halo in NGC~2613 will be discussed in
Sect.~\ref{no-lag}.

\subsection{The -307 km s$^{-1}$  Feature}
\label{307feature}

A distinct feature occurring at -307 km s$^{-1}$ (see
the Advancing panel
of Fig.~\ref{PV} and Fig.~\ref{channels})
has been mentioned in Sect.~\ref{disk-halo} and
extends to a (negative) {\it z} height  of 22 kpc. 
(This feature is also distinguishable on the major
axis PV slice of Paper I). 
The channel maps show a complex structure
below the plane, although
two spurs dominate.  
The feature is very well defined in
PV space and
contributions to the feature can be seen
in slices 6 through 9 which covers a range of 13 kpc in
projected distance along the disk of NGC~2613 
(Fig.~\ref{moments}a).
There is also a distinct
height ({\it z} $\sim$ 90$^{\prime\prime}$ = 11 kpc) 
at which the feature begins to
widen abruptly
in velocity (Fig.~\ref{PV}) and also (approximately) spatially 
(Fig.~\ref{channels}), 
a point we return to in the model of Sect.~\ref{buoyancy}. 
Further support for HI at high {\it z} at this
position comes
 from early 
 HI observations by Bottema (1989) which show an HI ``blob"
 extending to $\sim$ 25 kpc from the plane and
sharing the rotation of the underlying disk.  
In addition, our low
resolution 20 cm continuum map also shows a single large feature at
this position extending 21 kpc from midplane 
(Irwin et al. 1999).   Thus, we will treat the -307 km s$^{-1}$
feature to be a single, coherent structure, as implied by its
appearance in PV space, centered at a projected
galactocentric distance of 15.5 kpc.  
  
This feature occurs over
a restricted region of observed radial velocity 
($V_r$ = -307 $\pm$ 30 km s$^{-1}$).
Assuming that the feature originates in the
disk (see Sect.~\ref{very-high}),
this should allow
us to place limits on the position along the line of
sight from which it originates
 since the rotation curve of the galaxy is known.
The galactocentric radius is given by
$R\,=\,R_{proj}\,(V(R)/V_r)$
 (ignoring sin $i$ = 0.98)
where $R_{proj}$ is the projected galactocentric radius and
$V(R)$ is the circular velocity at $R$
which is approximately constant in this part of
the galaxy (Table 2). Allowing a $\pm$30 km s$^{-1}$
range on $V_r$, the feature could therefore have originated
from
 $R \,=\,R_{proj}\,=\, 15.5$
 kpc or anywhere within $\pm$ 7 kpc from this value along the line
of sight, corresponding to a range of $R$ between
15.5 and 17 kpc.  Since the projected size of the
feature is larger than this range (Fig.~\ref{channels}),
it is likely that there has been some expansion with
distance from the plane.

We can also put a limit on
the age of the feature.  Gas which leaves the disk
at an earlier time should have a component of line of sight
velocity corresponding to the motion of the underlying
disk at that time.  As the disk rotates, gas which leaves
later may have a different line of sight velocity component.
Consequently, there may be a gradual change in 
observed velocity with height (increasing or decreasing)
 depending on 
geometry.  Since we see no such change, we
can use the velocity half-width of the feature,
30 km s$^{-1}$, to
place a limit on the angle over which the
underlying disk has rotated over the age
of the feature. We find that
the feature must have
formed over a timescale which is less than the time
required for the disk
to rotate $\pm$ 24$^\circ$  at
galactocentric
radii between 15.5 and 17 kpc. From the circular
velocity in this region, the corresponding
timescale is 4.2 $\times$ 10$^7$ yr.  These arguments
will depend on how the ejected gas may or may
not be coupled to
the disk and should be considered order of magnitude only.
Nevertheless, the result is similar
to the kinematic ages found for expanding supershells
in other galaxies (cf. King \& Irwin 1997,
Lee \& Irwin 1997, Lee et al. 2001).  Basically the
``straightness'' of the feature in PV space requires
that the feature have an age $<<$ the rotation period.

The mass of this feature, as measured in PV space to
the level at which the feature blends with the disk, is 
($8\,\pm\,2)\,\,\times\,10^7$ M$_\odot$.  Approximating
the feature as a cylinder of height, 22 kpc, width and
depth, 13 kpc, the mean density is $1.1\,\times\,10^{-3}$ 
cm$^{-3}$.  (Reducing the diameter of the cylinder to
7.3 kpc (150$^{\prime\prime}$) results in a mean density
of $3.6\,\times\,10^{-3}$ cm$^{-3}$.)
If the feature has reached a {\it z} height of 22 kpc in
4.2 $\times$ 10$^7$ yr, the implied mean
outflow velocity is 512 km s$^{-1}$. The corresponding 
HI mass 
outflow rate would be 1.9 M$_\odot$ yr$^{-1}$.


We can also compute the potential energy of this feature
(see Lee et al. 2001) which requires a knowledge of
the mid-plane stellar density and scale height.
Comparing the rotation
curve of NGC~2613 (Chaves \& Irwin 2001) to that of the Milky Way
(Sofue \& Rubin 2001), NGC~2613 is a factor of 4 more
massive, and factor of 2 larger in radius.  NGC~2613 also has
a thinner disk, if we take the thickness of the HI disk 
(188 pc, Table 2)
to be representative of the stellar thickness. 
 Scaling
the mid-plane stellar density from the Galactic
value (0.185 M$_\odot$ pc$^{-3}$) 
yields 0.74 M$_\odot$ pc$^{-3}$ for NGC~2613
and
adopting
 100$^{\prime\prime}$ (12.6 kpc) (about half the full
{\it z} extent) to be representative
of the height of the feature above the plane, we find
a potential energy of 1.5 $\times$ 10$^{56}$ ergs
for
 the -307 km s$^{-1}$ feature.  Thus, energy of this
order is required to transport the cool gas to the heights
observed.
 Although very large, this
value is not unlike that determined for
the largest supershells in other edge-on galaxies
(cf. 
an input energy of $\sim$ 3 $\times$ 10$^{56}$ ergs for NGC~3556, 
King \& Irwin 1997).  We return to this issue in Sect.~\ref{buoyancy}.

\section{Discussion}
\label{discussion}

\subsection{The Absence of a Lagging HI Halo in NGC~2613}
\label{no-lag}

There are now several 
galaxies for which there is evidence for lagging halos.
 In NGC~891
(i $\ge$ 88.6$^\circ$), the 
rotation curve of the HI halo between {\it z} =
 1.4 to 2.8 kpc reveals 
a peak velocity $\sim$
25 km s$^{-1}$ lower than in the disk  (Swaters et al. 1997).
In NGC~5775 (i = 86$^\circ$), the ionized gas velocities
decrease with
{\it z} to heights of 5 - 6 kpc above which the velocity remains roughly
constant (Rand 2000).
The HI in this
galaxy shows more complex structure, though lagging gas tends to dominate
at {\it z} heights up to 5 to 6 kpc.
  Above this region, the velocities are more nearly constant
or the features break up into clumps which are seen over a
wide range in velocity
(Lee et al. 2001).  Models of NGC~2403 (inclination of 61$^\circ$)
suggest that a lagging halo extends to 
{\it z} $\sim$ 3 kpc (Schaap et al. 2000). 
Thus, from the sparse data available, 
the region of the lag extends to typically
3 to 6 kpc above the plane above which (if gas is detected at all)
there isn't strong evidence for a global lag. 
Given the presence of lagging halos in these other galaxies and
also some theoretical expectation of lags
(cf. Bregman 1980, Benjamin 2002,
Collins et al. 2002), the absence of a
lagging velocity gradient along the discrete features (Sect.~\ref{disk-halo})
and the absence of a global lag (Sect.~\ref{models}) in NGC~2613
require comment.

Firstly, given the inclination of this galaxy 
(79$^\circ$, Table 2) and its HI radius of $\sim$ 35 kpc,
the discrete features must reach a height of {\it z} $>$
6.5 kpc before they will be seen beyond
 the projection of the disk.
If it is true that lags only exist up to 3 to 6 kpc and above this
there is no longer a lag (see above), then lags (if they exist)
along the 
discrete features would be projected
against the background disk. This would make a velocity 
gradient (if present)  
impossible to detect along the discrete features.
The `straightness' of the -307 km s$^{-1}$ feature in
velocity (Fig.~\ref{PV}) to such high latitudes
beyond the projection of the disk is, however,
quite remarkable.  If this feature emerges into a
hot X-ray corona, as suggested in 
Sect.~\ref{buoyancy}, then either the
velocity of the broad-scale X-ray
corona is 
not appreciably different from the underlying
disk (although the velocity dispersion is likely much higher),
or the timescale over which it
could appreciably affect this gas column is greater than the age
of the feature.

 As for a globally lagging halo,
our kinematical model 
is indeed capable of detecting such a lag,
even if it is projected against the disk
(see Sect.~\ref{models}) provided
there is sufficient high latitude emission 
that halo gas can be detected at all.
What we have found, however, is that NGC~2613
does not have an HI thick disk or halo.  The
vertical exponential scale height (188 pc, Table 2)
indicates that the disk is, in fact, thin.  If we
create
a model galaxy using the parameters of Table 2,
it is straightforward to show 
that emission from all high latitude
({\it z} $>$ 1 kpc) HI in fact falls
below the map noise.  Thus, NGC~2613 does not have
a detectable lagging halo because it does not have
a halo at all.  

As a 
further comparative test, we consider the HI halo of 
another galaxy at the same distance, NGC~5775 (i = 86$^\circ$,
D = 25 Mpc), for
which we detected an HI halo with exponential scale
height of 9.14$^{\prime\prime}$ (1.1 kpc) using the
same technique
(Irwin 1994).  If we
create a model for the lower 
inclination   (i = 79$^\circ$)  NGC~2613 in which
the vertical density distribution declines exponentially
from its in-disk value
with the same scale height as
 NGC~5775,
we find that it could easily have been detected,
even at the lower inclination of NGC~2613.

This result illustrates the importance of using a model,
rather than PV slices alone, in 
drawing conclusions
about the possible existence of lagging halos.  For
example, the model can disentangle
inclination effects from the effects of a
thick disk.   For NGC~2613,
 since the modeled inclination
is now known to be on the low end of the
range quoted in Chaves \& Irwin (2000), our earlier conclusion of 
a lag with {\it z} can 
now be largely explained by projection against the background disk.
This is succintly illustrated via the dashed curves in 
Fig.~\ref{PV} which show that our thin disk model provides
an almost perfect fit to the data.


\subsection{The Discrete Features: Internal Origin and Energies}
\label{very-high}

We have
argued elsewhere (Paper I), largely on the basis of 
above/below plane symmetry (Fig.~\ref{Carray}),
that the observed features 
are internally generated and
represent outflows.   While these new data
show that the galaxy is indeed interacting, it is unlikely that most of
the observed features are produced via cloud impacts since the
cloud would have to pass completely through the disk, forming similar
structures  on both sides, in
contrast to what is expected theoretically 
(see Santill{\'a}n et al. 1999).
Anomalous
velocities are also more likely in the case of impacts, 
whereas the velocities observed in NGC~2613 are 
typical of the underlying disk.
Thus, the new data are consistent with the outflow interpretation. 
  We cannot, however,
rule out the possibility of some impacting clouds in this system 
and
it is also possible that the interaction may
assist in the process of disk-halo dynamics,
for example via stimulating a starburst or 
instabilities such as the Parker instability.

If the features are internally generated, their very large sizes
({\it z} up to 28 kpc, Sect. \ref{disk-halo} imply exceptional
energies ($\sim$ 10$^{56}$ ergs) which are difficult to conceive
of, especially at large distances from the nuclear vicinity and
in a galaxy which is not a starburst.  This energy
problem has been known since the first detections of Heiles
Shells in the Milky Way
(Heiles 1979, 1984) and has generated suggestions as far
ranging as multiple supernovae and stellar winds
(Heckman 2001),
gamma ray bursters
(Efremov et al. 1998, Loeb \& Perna 1998), and jet bubbles
(Gopal-Krishna \& Irwin 2000)
to help explain the high energies. 
 In the following section,
we suggest that the exceptional heights achieved by the
features in NGC~2613 may be
as much a result of the environment
into which the features emerge,
as to the input source itself.

\subsection{Buoyant Outflow}
\label{buoyancy}

The -307 km s$^{-1}$  feature 
(Fig.~\ref{PV}, Advancing) is remarkable in  
its well defined structure, in the very high latitude (22 kpc)
that it achieves, and in its sudden and dramatic widening
and break up in velocity 
at a height of $\sim$ 11 kpc. 
 In individual slices (see
slices 8 and 9, in particular), this velocity
widening gives the feature a mushroom-like appearance
and is reminiscent of
buoyant gas rising through a higher density
medium.  A precedent for this kind of behaviour is
the 350 pc Galactic Mushroom discovered in
Canadian Galactic Plane Survey data and which
has been interpreted in terms of buoyant
outflow (English et al. 2000).  

Given this similarity, we here consider whether  buoyant
outflow could explain the observed high latitude discrete 
HI features in NGC~2613. 
 We present the following as a 
feasibility study only to see whether the results provide
a reasonable match to the data using realistic parameters.
We consider a match to the -307 km s$^{-1}$  feature
only, at this time, since it is the clearest case amongst
the disk halo features in NGC~2613; however, if the model
is correct, it must clearly apply to the other features
as well.
A similar development has been presented by 
Avillez \& Mac Low (2001) for smaller
features like the Galactic Mushroom, but we here 
consider much larger scales (many kpc), and also include
the effects of drag.  

The scenario envisioned is one in which a hot, X-ray emitting
corona already exists around the galaxy, possibly set up via
venting through 
previous fountain or chimney activity.  Some event or events
occur within the disk which are sufficiently energetic to
produce blow-out.  The HI is already entrained
or swept up in some way by the time
the outflow emerges into the halo.  
An initial velocity (i.e. the velocity at
blow-out) could be present but is not included here.
Thus, the velocities achieved in the plume are
a result only of buoyant forces.  The attraction of
this model is that, rather than requiring that all of the
energy in the HI plume be supplied by the instigating
event in the disk ($10^{56}$ ergs, Sect.~\ref{307feature}),
we require only enough energy to produce the blow-out condition
(e.g. $10^{53}$ ergs, Tomisaka 1998).  The remaining
energy is extracted from the pressure gradient in the
hot coronal gas.  Ultimately, the galaxy's potential itself
is providing the energy source. 

Although we do not investigate the details of the entrainment
or sweeping up of
HI, we assume that the hot gas inside the plume carries the
cool HI with it and that the density of the HI
declines with {\it z}  in a fashion
 similar to the hot buoyant plume material.
In this feasibility study, we do not consider structure
in the corona or outflow plume, other than
the cylindrical geometry chosen for the plume, and
also neglect the effects of shocks. The integrity of the HI in
the presence of hot gas will be considered in the next section.

The stellar/mass density distribution of the thin
disk is described by:
$$\rho_*(z)\,=\,\rho_*(0)\,sech^2({z\over{2z_*}})$$ 
where $z_*$ is the scale height of the stellar disk, and
$\rho_*(0)$ is the stellar density at mid-plane.
Integration of Poisson's Equation:
$$ {{dg}\over{dz}}\,\,=\,\,-4\pi\,G\,\rho_*(z)$$
over the regime, $z\,>>\,z_*$, together with the above
density distribution, results in a gravitational
acceleration which is constant with $z$ and has a magnitude:
\be{g}
g\,=\, 8\,\pi\,G\,\rho_*(0)\,z_*   
\ee
The hot coronal gas (subscripted, $c$) is taken to be isothermal
at temperature, $T_c$, with
an exponential fall-off in both pressure and density, respectively:
\be{pressure_density}
P_c(z)\,=\,P_c(z_*)\,e^{-z/H}~~~~~~~~~~~~~~
\rho_c(z)\,=\,\rho_c(z_*)\,e^{-z/H}
\ee
where $H$ is the
scale height. This hot coronal distribution starts
at the top of the (HI + stellar)
thin disk, i.e. at $z_*$ where the outflow just
blows out of the thin layer.  
The scale height is given by 
\be{scale_height}
H\,=\,k\,T_c/(g\,\mu\,m_p)
\ee
where 
$k$ is Boltzmann's constant, $\mu$ is the mean molecular
weight and $m_p$
is the mass of the proton.  We consider a pure
hydrogen gas  ($\mu$ = 1) for simplicity. 

Inside the plume (subscripted, $i$), we consider the gas
to be adiabatic.  The temperature gradient
is:
\be{adiabatic}
{{dT_i}\over{dz}}\,=\,(1-{1\over\gamma})\,
{{T_i}\over{P_i}}\,{{dP_i}\over{dz}}
\ee
For a plume in pressure equilibrium with
its surroundings,
\be{pressure_interior}
P_i(z)\,=\,P_c(z_*)\,e^{-z/H}
\ee
where, again, we take the plume base to occur at $z_*$.
Substituting  Eqn.~\EC{pressure_interior}
 and its derivative 
into Eqn.~\EC{adiabatic} yields:
\be{temperature_interior}
T_i(z)\,=\,T_i(z_*)\,e^{({{{(1-\gamma)}\over\gamma}}{z\over H})}
\ee
and from Eqn.~\EC{pressure_interior}, Eqn.~\EC{temperature_interior}
and
the perfect gas law:
\be{density_interior}
\rho_i(z)\,=\,\rho_i(z_*)\,e^{ -( { {z}\over{\gamma H} } )     }
\ee

The equation of motion (force per unit volume) of the plume material is:
\be{force_equation}
\rho_i(z)\,{{dv}\over{dt}} \,=\, \rho_c(z)\, g\,-\, \rho_i(z)\,g \,-\,
C_D\rho_c(z)\,{v^2\over z} \,-\,
{{2\,\eta\,v}\over{r^2}}
\ee
The first term on the right
hand side denotes the buoyancy force, the second term gives the
weight of the plume material, the third term specifies the
drag force against the upper surface (assuming
a case in which the effects of this
force can propagate through the column) 
where $C_D$ is the dimensionless drag coefficient,
and the fourth term
represents the viscous drag on the cylinder sides, where
r is the cylinder
radius
 and $\eta$ is the 
coefficient of viscosity ($g\,cm^{-1}\,s^{-1}$). 

Both
$C_D$ and $\eta$
depend on the Reynolds number:
$$
Re\,\approx\,
5\,\times\,10^{14}\,\, {{\rho_c(z)\,L\,v\,ln\Lambda}
\over
{T_c^{5/2}}}
$$
assuming a pure ionized hydrogen gas, where L is a scale length
above which motion is damped by viscous effects and
$$\Lambda\,\approx\,
1.3\,\times\,10^4\,\,
{{T_c^{3/2}}
\over
{n_e(z)^{1/2}}}
$$
where $n_e(z)$ is the coronal electron density.  Taking
$v = 500$ km s$^{-1}$ (Sect.~\ref{307feature}) and $T_c\,=\,
2\,\times\,10^6$ K, we find 
for size scales, $L$, of order several
kpc, $Re$ is in the range $10^{3\,\to\,4}$. In this range of
$Re$, 
$C_D\,\sim\,1$ for an incompressible fluid with cylindrical geometry.
The viscosity coefficient
includes both turbulent and molecular
terms, i.e. $\eta\,=\,\eta_t\,+\,\eta_m$, where
 $\eta_m$ = $L\,v\,\rho_c(z)/Re$ and 
$\eta_t \,=\,(Re/Re_{crit})\,\eta_m$, where $Re_{crit}$
is the critical Reynolds number which designates
the value of $Re$ at which the flow becomes
turbulent. $Re/Re_{crit}$ is not known
and depends on the geometry and nature of the
interface,
but typically has values between 1 and 100.
We can consider whether the drag force
on the cylinder sides
(4th term of Eqn.~\EC{force_equation})
 is appreciable in comparison to
the drag on the top of the cylinder 
 (3rd term of Eqn.~\EC{force_equation}).
Taking $L\,\sim\,r$,
the
ratio of the 3rd to 4th drag terms becomes
[$(Re/2)\,(r/z)\,1/(1\,+\,Re/Re_{crit})$].
Since $r$ is of order $z$, then for
 $Re\,\sim\,10^{3\,\to\,4}$, even  
$Re/Re_{crit}$ up to 100 ensures that
the drag at the top will dominate over that
at the sides.
Therefore the 4th term in
 Eqn.~\EC{force_equation} is small
and will be neglected.

Substituting  $v = dz/dt$, ${\cal{R}}(z_*)\,=\,\rho_c(z_*)/\rho_i(z_*)$,
and $\gamma\,=\,5/3$ into
Eqn.~\EC{force_equation}
and rearranging yields:

\be{equation_of_motion}
{{dv}\over {dz}}\,=\, {\cal{R}}(z_*)
\,e^{   -0.4{z\over H}   }
\,{g\over v}\,-\,
{g\over v}\,-\,
{\cal{R}}(z_*)\,
e^{-0.4{z\over H}   }
\,{{v}\over z}
\ee

This equation was integrated numerically for the
input parameters shown in Table 3, yielding
the curves of $v(z)$ and $z(t)$ shown in
Fig.~\ref{vz} and Fig.~\ref{zt}, 
respectively.  The stellar scale height,
$z_*$ is set to 188 pc (Table 2) and
we consider a coronal temperature, $T_c\,=\,2.5\,\times\,10^6$ K
(Models 1 to 4)
which is comparable to that found from X-ray observations of
NGC~253 (Pietsch et al. 2000) as well as a temperature which is
a factor of 2 higher (Model 5).
Since the midplane stellar density of 0.74 M$_\odot$ pc$^{-3}$
is a value which has been scaled from
Galactic values, given the size and mass of NGC~2613
(see Sect.~\ref{307feature}), we also consider a slightly
lower value (Models 3 and 4).  The behaviour of the curves
depends only on the density ratio at the base of
the corona, ${\cal{R}}(z_*)$, rather than the individual
densities, but we can additionally fix 
the density at the bottom of the corona,
$n_c(z_*)$ to be
equal to the HI density at the top of the thin disk 
at the position of the plume (see Table 2), providing
constraints upon the density within and outside of the plume.
The peaks of the velocity curves 
(Fig.~\ref{vz}) indicate where the
acceleration of the plume material goes to zero; we define the
$z$ height at which this occurs to be $z_{stall}$.  This
position should
correspond to the point at which the plume widens in velocity
space, observationally determined to be $\sim$ 11 kpc.  

Models 1 and 2 
(Fig.~\ref{vz}) show the effect of changing the initial density
ratio, such that the lower ratio (Model 2) results in a lower maximum
velocity and a lower $z_{stall}$ (7.4 kpc as compared to 11.9 kpc).
A comparison of Models 1 and 3 or of Models 2 and 4 show the
effect of decreasing the midplane mass density.  This lowers the
gravitational acceleration (Eqn.~\EC{g}) which lowers both the
buoyancy and the weight of the plume material.  It also increases
the scale
height (Eqn.~\EC{scale_height}) which increases both the buoyancy
and the drag.  The net effect is higher $z_{stall}$ at lower $g$
for the range of parameters given here.  A comparison of Models 1
and 5 shows the effect of increasing the coronal gas temperature.
The higher temperature increases the scale height alone
(Eqn.~\EC{scale_height}) which, again, increases both 
buoyancy and drag.  Since drag is velocity dependent, the
net effect is that the peak velocities achieved are lower for
higher coronal temperatures. 

Clearly, Models 1, 4, and 5 provide an adequate match to the
observed $z_{stall}$.  Model 1, however, results in a mean plume
density (between $z_*$ and
$z_{stall}$) of only $0.27\,\times\,10^{-3}$ cm$^{-3}$ whereas 
the HI density alone is $\sim$ $1\,\times\,10^{-3}$
(Sect.~\ref{307feature}), and is therefore not
realistic.   Models 4 and 5 both result in reasonable
fits to the known observational parameters, with Model 5
slightly preferred because of its higher internal plume density
(higher than that of the HI alone)
and lower internal temperature.  Note that the mean internal temperature
derived here
($\sim\,10^7$ K) is comparable to the hot (T = 1.2 keV = 
$1.4\,\times\,10^7$ K) component of  the outflow
in NGC~253 (Pietsch et al. 2000).
The peak velocities derived here ($\sim$ 250 to 300
km s$^{-1}$) are lower than the gross estimate of 500
km s$^{-1}$
computed in Sect.~\ref{307feature} since we only model  the
feature up to the stall height (11 kpc) rather than over
its total length.  The timescales of $\sim$ 5 to 6 $\times$
10$^7$ yr 
(Fig.~\ref{zt}) are in good agreement with the estimate
found from the kinematical structure of the plume 
(4$\,\times\,$10$^7$ yrs).  The mass outflow rate to
the stall point is $\sim$ 1 M$_\odot$ yr$^{-1}$. 
 Mass flow continuity
requires that the  plume of Model 5 
should increase in radius by a factor of 1.8 between
{\it z} heights of 4.7 and 11.1 kpc.

\subsection{Integrity of the HI}

 It is unlikely that the HI will be
uniformly distributed at its mean density of $\sim$ 
$10^{-3}$ cm$^{-3}$ but may exist in denser clouds or
clumps whose sizes and distribution are not
described by this simple model.  In general,
however, the situation
will be not unlike
that of those high velocity clouds (HVCs) which are in the
hot halo of the Milky Way 
and the same issues
regarding whether or not the clouds can remain neutral
must be considered.
 Photoionization of the HI by
starlight should be negligible, considering the
high galactic latitudes achieved. As pointed out
by Murali (2000), the relevant interactions are
HI - proton and HI - electron interactions for
which the interaction cross-sections are in the
range $\sigma\,=\,10^{-16}$ to $10^{-15}$ cm$^{-2}$
for a relative velocity of order 200 km s$^{-1}$ as
is appropriate here.  The mean free path into the
HI cloud is then l = 1/($n_c \sigma$) = 0.3 to 3 pc
or smaller if the HI is clumped.  This is considerably
smaller than the size scale of the plume and
therefore these ionizing interactions should be minor.


The most important interaction will be
heating due to thermal conductivity leading to
the evaporation of clouds.
  The classical
mass evaporation rate,  applicable to 
 the case in which the mean free path is small
in comparison to cloud size,
 is given, for approximately spherical clouds, by
$\dot m$ = 2.75 $\times$ 10$^4$ T$^{5/2}$ R$_{pc}$ $\phi$ g s$^{-1}$,
where T is the temperature of the external medium, 
R$_{pc}$ is 1/2 of the largest dimension of the cloud
in parsecs,
and $\phi$ is a parameter which measures the
inhibition of heat flux due to the magnetic field
and cloud geometry (Cowie \& McKee 1979, Cowie et al. 1981).
 We will assume 
$\phi$ = 1 (no inhibition)
 which maximizes the evaporation rate.
Using T = 2.5 $\times$ 10$^6$ K and R = 5 $\times$ 10$^3$ pc
we find $\dot m$ = 0.021 M$_{\odot}$ yr$^{-1}$
(to within factors of a few, given the different
geometry).  The timescale for
complete evaporation of the 8$\,\times\,$10$^7$
M$_{\odot}$ HI plume  at a constant rate is then
M/ $\dot m$ = 4 $\times$ 10$^9$ yr.  Since
this is two orders
of magnitude larger
 than the age of
the plume, we expect that the plume will
not evaporate over its lifetime.   A caveat, however,
is 
that since
 M/ $\dot m$ $\propto$ R$^2$, if HI is distributed
in many smaller clouds, then the
evaporation time could approach the age
of the plume.

\subsection{The X-ray Corona}
\label{x_ray_halo}

If the HI is indeed reaching such high
values of {\it z} because of buoyancy, then 
 HI disk-halo features probe the
parameters of the halo gas,
as suggested in Sect.~\ref{introduction}.  We have so
far
considered only whether buoyancy is
feasible.  There may be other dynamics
at work as well (for example, an initial
velocity at blow-out or magnetic fields) and the proposed
hot corona is also unlikely to have the
smooth distributions postulated
here.  Nevertheless, in the
context of the model, it is interesting
to predict the X-ray luminosity of the corona
in the vicinity of the plume.

For the case in which the
 ion and electron
densities are equal, the index
of refraction is 1, the charge  $Z$ =1,
 the frequency is greater
than the plasma frequency, and 
$T\,>\,3.6\,\times\,10^5$ K, then:
$$
L_x\,=\,3.7\,\times\,10^{-38}\,V\, 
{{n_e^2}\over{T^{1/2}}}
\int
ln[4.7\,\times\,10^{10}\,(T/\nu)]
\,exp(-h\nu/kT)d\nu~~~~~~~~~~
{\mathrm{ergs\,\, s^{-1}}}
$$
 (Lang 1999), where $V$ is the source volume and
$n_e$ is the electron density.
Taking the limits of integration to be over the soft X-ray
band (0.2 to 2.5 keV) and $T_c\,=\,5\,\times\,10^6$ K,
(Model 5),
the above becomes:
$$L_x\,=\,2\,\times\,10^{36}\,
({V\over{\mathrm{kpc}^3}})\,
{({n_e\over{0.01\,{\mathrm{cm^{-3}}}}})}^2~~~~~~~~~~
{\mathrm {ergs\,\,s^{-1}}}
$$
A mean coronal density (between $z_*$ and
$z_{stall}$) of 
 0.035
$\mathrm{cm^{-3}}$ (Model 5, Table 3) in a 
rectangular region roughly the size of the
modeled region of the 
plume (11 kpc $\times$ 13$^2$ kpc) provides
an estimate of the
soft X-ray luminosity of $L_x$ =
$4.6\,\times\,10^{40}$ erg s$^{-1}$. Using
instead
the density at $z_{stall}$ of
 2.8 $\times$ 10$^{-3}$ 
$\mathrm{cm^{-3}}$ gives
$L_x$ = $2.9\,\times\,10^{38}$ erg s$^{-1}$.
The
X-ray luminosities of the hot gas in NGC~253
and M~82 are
$2\,\times\,10^{39}$ erg s$^{-1}$ and
$2\,\times\,10^{40}$ erg s$^{-1}$, respectively
(Pietsch et al. 2000).  Thus, the predicted
X-ray halo for NGC~2613 should be observable
below height, $z_{stall}$,
though is unlikely to have the smooth exponential
distribution envisioned in this simple model.
 Note that
NGC~2613 is  
more than an order of magnitude more massive
than M~82 
and therefore is more likely to retain any
hot gas which might be around it.  The values
computed here  
will be lower, of course, if some other effect
 is 
 contributing
to the outflow.  

\subsection{Blowout of Disk Gas}

The results of Table 3 show that it is feasible
to transfer large masses to
high galactic latitudes
via buoyancy in a postulated X-ray corona.  This 
drastically reduces the computed input energy
requirements since it is no longer necessary to eject
large masses to high altitudes.  It is only
necessary to 
 achieve blow-out
through the thin HI disk of NGC~2613.  The
conditions required for blow-out have been investigated
by a variety of authors (see Tomisaka 1998, for example)
but energy requirements are typically of order 
10$^{53}$ ergs, rather than the 10$^{56}$ ergs
that would normally be required for the
-307 km s$^{-1}$ feature.

While the details of 
the interaction between the HI and hot outflowing
gas are beyond the scope of this paper, it is
important to ask whether 
there would originally
have been sufficient HI in the disk from which
the 8 $\times$ 10$^7$ M$_\odot$ in the
-307 km s$^{-1}$ feature
 could have been swept up.
Assuming cylinderical geometry,
a disk region of radius, 6.5 kpc 
(Sect.~\ref{307feature}) and using the
modeled
density and thickness of the disk (Table 2),
the available HI mass is $\sim$ 3 $\times$ 10$^8$
 M$_\odot$ suggesting that $\sim$ 26\% of the
disk mass is swept up.  This fraction would be
higher for cone-like outflow.  

We should also consider whether
the mass outflow rate is 
consistent with that of a galactic fountain.
The estimated HI
mass outflow rate (Sect.~\ref{307feature},
Sect.~\ref{buoyancy}) is $\sim$ 1 to 2  
M$_\odot$ yr$^{-1}$.  Thus, the combined
HI + hot gas outflow rate will be of
order several 
M$_\odot$ yr$^{-1}$.
Collins et al. (2002) have estimated the
 mass
flow rate for the diffuse ionized 
gas (DIG) component in a 
 galactic fountain, assuming a
ballistic model of gas clouds.  They find
global values of 
$\dot M$ = 22 $\sqrt{f_V/0.2}$ M$_\odot$ yr$^{-1}$
for NGC~891 and 
$\dot M$ = 13 $\sqrt{f_V/0.2}$ M$_\odot$ yr$^{-1}$
for NGC~5775, where $f_V$ is the filling factor.
If these global values apply to a 20 kpc radius
disk, taking $f_V$ = 0.2 and
 scaling to the disk area of the 
-370 km s$^{-1}$ feature yields mass
outflow rates  of
1.4 and 2.3 M$_\odot$ yr$^{-1}$ 
which are
comparable to what we estimate, above.
It is not yet clear, however, whether these galactic
fountain values can be directly scaled to
the relevant regions of NGC~2613.  It may
be that  
some additional source of pressure is still
required,
for example, magnetic fields in the form of a Parker 
instability.  If such fields continue to rise
into the corona, the inclusion of this magnetic
pressure would relax (i.e. lower) the density 
or temperature requirements internal to the plume
(Table 3).

\section{Conclusions}
\label{conclusions}

New VLA D array data of NGC~2613 have been combined
 with previous higher resolution observations 
(Chaves \& Irwin 2001) to show
a more extensive HI distribution than previously observed.
The galaxy is now seen to have a tidal tail on its eastern
side due to an interaction with its companion,   
ESO~495-G017, to the north-west.  

The three-dimensional
HI distribution in NGC~2613 has been modeled
following Irwin \& Seaquist (1991) and Irwin (1994), a method
which allows the volume density distribution to be determined
as well as the scale height and inclination to be 
disentangled.  We find that the inclination of the galaxy
(79$^\circ$) is on the low end of the range given 
in Chaves \& Irwin (2001)
and the model now shows that there is no 
HI halo in NGC~2613.  Rather, the
 global HI distribution is well fit by a thin
disk of exponential scale height, $\it z_e$ = 188 pc.
The use of such a model is very important in
drawing conclusions about the presence or absence
of a global halo in a galaxy of this inclination.
Previous reports of a lagging halo from PV slices
alone can largely be
attributed to projection against the background disk. 

While there is no significant
global HI halo in
NGC~2613, there are  more
discrete  disk-halo HI features  
than previously detected and these HI features achieve 
extremely high latitudes.  Even though a tidal interaction is
occurring, we suggest that most of the discrete 
kpc-scale features have
been produced internally rather
than from impacting clouds, although we do not rule out
the possiblity of the companion galaxy having some
indirect
effect (e.g. triggering instabilities).
The presence of many discrete
features may be related to the fact that the global
HI disk is thin, favouring blow-out. The
observed  {\it z} heights are quite remarkable (e.g. up to
28 kpc).
The -307 km s$^{-1}$ feature, in particular,
below the plane  on the advancing side,
 reaching
22 kpc in $\it z$ height and of total mass, 
(8 $\pm$ 2) $\times$ 10$^7$ M$_\odot$,
is very obvious and well-defined in PV space.
Its center is likely close to its projected radius of 15.5 kpc
and it extends over a large ($\pm$ 7 kpc radius) projected
galactocentric radius. 
If this feature has achieved its
$\it z$ height as a result of internal processes, then extremely
large energies are required, $\sim$ 10$^{56}$ ergs.

Given the very high input energies required for the 
-307 km s$^{-1}$ feature, its resemblance to smaller
buoyant features (cf. the Galactic Mushroom,
English et al. 2000),
and the fact that X-ray halos are being found around an
increasing number of star forming spiral galaxies, we have
carried
out a feasibility study as to whether this feature can
be interpreted as an adiabatic buoyant plume. 
The observed HI would be carried out by a hot, low density
outflowing
gas and, after having achieved blowout, would
rise through a hot pre-existing X-ray corona.
A reasonable example (Model 5), 
has a mean plume temperature
and density of  $1\,\times\,10^7$ K and 
5.5 $\times$ 10$^{-3}$ cm$^{-3}$ rising into a hot
isothermal corona of
 temperature and
mean density,
$5\,\times\,10^6$ K and 0.035 cm$^{-3}$, respectively.
These conditions produce a stall height of 11 kpc which is
where the observed plume widens in velocity and position space. 
The coronal density at the stall height is 2.8 $\times$
10$^{-3}$ cm$^{-3}$.
The maximum outflow velocity 
in this model is 290 km s$^{-1}$ and it reaches
the stall height in 5.4$\,\times\,$10$^7$ yrs.  This model
shows that, even with buoyancy alone
(and there may be additional sources of pressure such 
as magnetic fields), HI can reach these
extreme {\it z} heights.

 The advantage
of such a model is that the energy requirements from
the initial event are
drastically reduced to being only what is required for
blowout,
a reduction of several orders of magnitude.  The energy
is largely extracted from the gravitational potential of the galaxy
rather than the initial event within the disk.  The behaviour
of the plume should sample the parameters of the 
X-ray corona.  
The predicted  X-ray luminosity suggests that the corona
should be observable at heights below the stall height
although we expect that the distribution of X-ray emission
may not be as smooth as assumed by the model.

\acknowledgments
JI wishes to thank the Natural Sciences and Engineering Research
Council of Canada for a research grant. 
 We are grateful to
Dr. R. N. Henriksen for fruitful and envigorating discussions.
Thanks also to Mustapha Ishak for assistance with MAPLE.

\clearpage

\begin{figure}
\caption{VLA C array HI Column density map of NGC~2613 superimposed
on a Digitized Sky Survey optical image, from Paper I.  Extensions
are marked F1 through F6.
\label{Carray}}
\end{figure}

\clearpage

\begin{figure}
\caption{(a) HI total intensity 
(zeroth moment) map of NGC~2613
from the combined C+D array data, rotated so that
the major axis is horizontal.  
The central positions of
NGC~2613 and its companion, ESO~495-G017, are marked with crosses.
The map was formed by
first smoothing the 
 data by 3 channels in velocity
and 11 pixels ($\sim$ 1.4 beams, 1 pixel = 
5$^{\prime\prime}$) spatially, applying a
cut off of 0.35 mJy beam$^{-1}$ to the smoothed, lower
rms cube, and then summing the unsmoothed
data excluding those points that fell below
the cutoff.
The greyscale ranges from 0 to 3000 
Jy beam$^{-1}$ m s$^{-1}$
and contours
are at 20, 30, 50, 90, 150,
250, 500, 1200, 2150, 2800 and 3200
Jy beam$^{-1}$ m s$^{-1}$
(1 Jy beam$^{-1}$ m s$^{-1}$ = 7.375 $\times$ 10$^{17}$ cm$^{-2}$).
Vertical numbered
lines indicate where the slices were taken for the PV
plots of Fig. 3 and the features, F1 to F6 are again labelled.
(b) The intensity-weighted mean velocity field (first
moment map).  Contours, in units
of km s$^{-1}$
are marked.  This map was made in the same fashion
as (a) but the cutoff was set to
0.45 mJy beam$^{-1}$.  
\label{moments}}
\end{figure}

\clearpage

\begin{figure}
\caption{Position-velocity plots corresponding to the
numbered
vertical slices shown in Fig. 2.  Solid curves 
represent the data and red dashed curves represent the
best fit model (see Table 2).  The velocity is
with respect to the systemic velocity.
Contours are at 0.65 (1.5$\sigma$), 0.85, 1.1,
1.5, 3, 6, 10, 15, and 22 mJy beam$^{-1}$.
Negative contours (not shown) are randomly
distributed over the images.  The
last two panels show an average over 250$^{\prime\prime}$
for the receding and advancing 
sides, respectively.
For these, contours are at 0.2, 0.35, 0.6, 0.9, 2, 3, 4.5,
6, 10, and 15 mJy beam$^{-1}$ and the rms noise is
 $\sim$ 0.16 mJy beam$^{-1}$.
\label{PV}}
\end{figure}

\clearpage

\begin{figure}
\caption{
Sequence of 4 channels on the advancing side of the
galaxy showing the -307 km/s feature with
contours and greyscale chosen to highlight this
feature.   Contours
are at -0.9, 0.9 (2$\sigma$), 1.2, 1.5, 1.9, 3, 6, 
and 10 mJy beam$^{-1}$ and the greyscale ranges
from 0.45 (1$\sigma$) to 10 mJy beam$^{-1}$.
Velocities are marked at the top right of each
panel.
\label{channels}}
\end{figure}

\clearpage

\begin{figure}
\caption{Zeroth moment map of the residuals
(data cube - best fit  model cube) l
for all points $>$ 3$\sigma$. Solid
curves represent excess data and dashed curves represent
excess model, with contours at
-800, -600, -400, -200, -100, 100, 200, and 400
Jy beam$^{-1}$ m s$^{-1}$.
\label{residuals}}
\end{figure}

\clearpage

\begin{figure}
\caption{Velocity as a function of {\it z} height for
the five buoyancy models, Model 1 (crosses), Model 2
(diamonds), Model 3 (circles), Model 4 (boxes) and 
Model 5 (points), whose parameters are given in Table 3.
\label{vz}}
\end{figure}
\clearpage

\begin{figure}
\caption{$z$ height as a function of time for
the five buoyancy models, Model 1 (crosses), Model 2
(diamonds), Model 3 (circles), Model 4 (boxes) and 
Model 5 (points), whose parameters are given in Table 3.
The integration of these curves have been halted at
$z_{stall}$.
\label{zt}}
\end{figure}

\clearpage

\clearpage 

\begin{deluxetable}{lcc}
\tabletypesize{\scriptsize}
\tablecaption{Observing and Map Parameters. \label{tbl-1}}
\tablewidth{0pt}
\tablehead{
\colhead{Parameter} 
& \colhead{D Array}
& \colhead{C+D Array} 
}
\startdata
No. velocity channels & 63 & 63 \\
Velocity Resolution (km s$^{-1}$) & 20.84 & 20.84 \\ 
Total Bandwidth (MHz) & 6.25 & 6.25 \\
Synthesized Beam & & \\
~~~major $\times$ minor axis 
($^{\prime\prime}$ $\times$ $^{\prime\prime}$) @ PA ($^\circ$) &
96.5 $\times$ 42.9 @ -21.6 & 47.1 $\times$ 32.1 @ -8.2 \\
Rms noise/channel (mJy beam$^{-1}$) & 0.93 & 0.45 \\
Rms noise/channel (K) & 0.14 & 0.18 \\
 \enddata




\end{deluxetable}


\clearpage

\begin{table}
\begin{center}
\caption{Modeled Kinematic and Density Parameters for Combined C+D Array Data.\label{tbl-2}}
\begin{tabular}{|l|c|}
\tableline\tableline
Parameter  & Whole Galaxy$^a$ \\
\tableline
RA (J2000) (h m s) &   08 33 22.8 $\pm$ 0.3  \\
DEC (J2000) ($^\circ$ $^\prime$ $^{\prime\prime}$)   
&   -22 58 29  $^{+4}_{-2}$ \\
PA ($^\circ$) & 114.2 $^{+0.8}_{-0.1}$ \\
i ($^\circ$) &  79.2 $^{+0.7}_{-0.1}$  \\
V$_{sys}$ (km s$^{-1}$) & 1663 $^{+8}_{-1}$  \\
V$_{max}$ (km s$^{-1}$) & 304 $\pm$ 4   \\
R$_{max}$ ($^{\prime\prime}$)  & 112 $^{+1}_{-59}$  \\
m & 0.51 $^{+0.05}_{-0.35}$   \\
$\sigma_V$ (km s$^{-1}$)&  17 $\pm$ 5   \\
n$_0$ (cm$^{-3}$)$^b$&   0.43 $^{+0.01}_{-0.26}$  \\
R$_0$ ($^{\prime\prime}$)&  140 $\pm$ 5  \\
D$_o$ ($^{\prime\prime}$) & 95 $^{+1}_{-5}$   \\
D$_i$ ($^{\prime\prime}$) & 81 $^{+15}_{-3}$   \\
H$_e$ ($^{\prime\prime}$) & 1.5 $^{+2.0}_{-0.1}$   \\
\tableline
\end{tabular}


\tablenotetext{a}{Errors reflect the variation in the
parameter that results when the C array and D array data
are modeled separately, and so throughout.}
\tablenotetext{b}{The density is the mean over the
physical region corresponding to a 5$^{\prime\prime}$ by
5$^{\prime\prime}$ by 20.84 km s$^{-1}$ (RA, DEC, Velocity)
voxel.
}
\end{center}
\end{table}

\clearpage

\begin{table}
\begin{center}
\caption{Buoyancy Model Parameters.\label{tbl-3}}
\begin{tabular}{|l|c|c|c|c|c|}
\tableline\tableline
Parameter & Model 1 & Model 2 & Model 3 & Model 4 & Model 5 \\
\tableline
 $T_c$ ($10^6$ K) & 2.5 & 2.5 & 2.5 & 2.5 & 5.0 \\
$\rho_*(0)$ ($M_\odot\,pc^{-3}$) & 0.74 & 0.74 & 0.50 & 0.50 & 0.74 \\
H$^a (kpc)$ & 1.4 & 1.4 & 2.1 & 2.1 & 2.8 \\
${\cal{R}}(z_*)$ & 100 & 20 & 100 & 20 & 10 \\
$n_c(z_*)\,^b$ (cm$^{-3}$) & 0.15 & 0.15 & 0.15 & 0.15 & 0.15 \\
$n_i(z_*)$ (cm$^{-3}$) & 0.0015 & 0.0075 & 0.0015 & 0.0075 & 0.015 \\
\tableline
$z_{stall}$ (kpc) & 11.9 & 7.4 & 17.5 & 11.0 & 11.1 \\
$\overline{n_c}\,^c$ (cm$^{-3}$) & 0.015 &
0.025 & 0.016 & 
0.026 & 0.035 \\
$n_c(z_{stall})\,^d$ ($10^{-3}$ cm$^{-3}$) & 0.031 & 0.76
  &0.036 & 0.80
  & 2.8  \\
$\overline{n_i}\,^e$ ($10^{-3}$ cm$^{-3}$) & 0.27 & 2.1
 & 0.28 & 2.2
&  5.5 \\
$\overline{T_i}\,^f$ ($10^7$ K) & 6.8 & 2.0 & 6.9 & 2.0 & 1.2 \\
\tableline
\end{tabular}
\tablecomments{ See Sect.~\ref{buoyancy} for an explanation of
the parameters.
$a)$ computed from Eqn.~\EC{scale_height} and 
Eqn.~\EC{g} using
$z_*\,=\,188$ pc (Table 2).
$b)$ The coronal density at
 $z_*$ (the top of the thin disk) is computed from the parameters
of the HI distribution at the position of the plume (see Table 2). 
$c)$ The mean coronal density is determined from Eqn.~\EC{pressure_density},
over the region
from $z_*$ to $z_{stall}$.
$d)$ The coronal density at height, $z_{stall}$ from 
Eqn.~\EC{pressure_density}.
$e)$ The mean plume density is determined from Eqn.~\EC{density_interior},
over the region
from $z_*$ to $z_{stall}$.
$f)$ The mean plume temperature is determined from Eqn.~\EC{temperature_interior},
over the region
from $z_*$ to $z_{stall}$.
}

\end{center}
\end{table}






\end{document}